\documentclass[aps,prd,english,superscriptaddress, longbibliography,11pt,notitlepage, nofootinbib]{revtex4}

\usepackage[colorlinks=true, a4paper=true, pdfstartview=FitV,
linkcolor=blue, citecolor=blue, urlcolor=blue]{hyperref}

\usepackage{amsmath}
\usepackage{amssymb}
\usepackage{graphicx}
\usepackage{hhline}
\usepackage{mwe}
\usepackage{amsbsy}
\usepackage{textcomp}
\usepackage{commath}
\usepackage{subfig}
\usepackage{slashed}
\usepackage{natbib}
\usepackage{xcolor}

\newcommand{\be}{\begin{equation}}
\newcommand{\ee}{\end{equation}}  
\newcommand{\bea}{\begin{eqnarray}}
\newcommand{\eea}{\end{eqnarray}} 
\newcommand{\half}{\frac{1}{2}}
\newcommand{\eps}{\epsilon}
\newcommand{\pr}{\partial}

\newcommand{\PhiN}{\Phi^{\{N\}}}
\newcommand{\intd}{\int_0^{2\pi}}
\DeclareMathOperator{\sech}{sech}

\usepackage{stackengine}
\stackMath

\makeatletter
\usepackage{babel}
\numberwithin{equation}{section}
\begin{document}
\immediate\write16{<<WARNING: LINEDRAW macros work with emTeX-dvivers
                    and other drivers supporting emTeX \special's
                    (dviscr, dvihplj, dvidot, dvips, dviwin, etc.) >>}

%\title{Residual Structure of the Fodor Expansion for Oscillons in
%$1+1$ Dimensions}
\title{\Large{Momentum Flux for Small-Amplitude Oscillons}}
                  
\author{\bf{N.~S. Manton}}
\email[]{N.S.Manton@damtp.cam.ac.uk}
\affiliation{Department of Applied Mathematics and Theoretical Physics,
University of Cambridge, Wilberforce Road, Cambridge CB3 0WA, U.K.}

\author{\bf{A. Mohammadi}}
\email[]{azadeh.mohammadi@ufpe.br}
\affiliation{Departamento de Física, Universidade Federal da
Pernambuco, Av. Prof. Moraes Rego, 1235, Recife - PE - 50670-901,
Brazil}

\begin{abstract}
An oscillon in scalar field theory in $(1+1)$-dimensions is an
oscillating, localised, almost periodic classical solution that slowly
radiates and decays to the vacuum solution. We calculate its momentum
flux integrated over a period, which would be zero for an exact
breather. For an oscillon of small amplitude $\eps$, truncated at
order $\eps^N$, we show that the integrated momentum flux is of leading order
$\eps^{N+4}$ for even $N$ and $\eps^{N+5}$ for odd $N$,
except for $N=1$, where it is of order $\eps^4$. It is therefore
very small.

\end{abstract}

\maketitle

%%%%%%%%%%%%%%%%%%%%%%%%%%%%%%
\section{Introduction}
\label{intro}

Consider a real scalar field $\phi(t,x)$ in $(1+1)$-dimensions, with
Lagrangian density\footnote{Throughout, subscripts $t,x,\tau,\zeta$
(usually without commas) denote partial derivatives. For functions
of a single variable, derivatives are sometimes denoted by primes.} 
\be
{\cal L} = \half \phi^2_t - \half \phi^2_x - V(\phi) \,.
\ee
$\phi$ satisfies the field equation
\begin{equation}
\phi_{tt}-\phi_{xx}+V'(\phi)=0 \,.
\label{eq:field_equation}
\end{equation}
We take $\phi=0$ to be a vacuum and $V$ to have the general analytic
form
\begin{equation}
V(\phi)=
\frac{1}{2}\phi^2 +\sum_{j=2}^{\infty}\frac{g_j}{j+1}\phi^{j+1} \,.
\label{eq:potential}
\end{equation}
The quadratic term in $V$ has unit normalisation, so the mass
threshold is 1 and the frequency of linearised waves is at least 1.
The couplings (coupling constants) $g_j$ are assumed to have generic
values, except where explicitly stated otherwise.

The momentum density and momentum flux, i.e. the components of
the energy-momentum tensor
\be
T_{\{xt\}} = -\phi_t \phi_x \,, \quad T_{\{xx\}} = \half \phi^2_t
+ \half \phi^2_x - V(\phi) \,,
\ee
obey the current conservation equation $T_{\{xt\}t} + T_{\{xx\}x} = 0$.
Integrating over a space-time rectangle $[0,T]\times[0,X]$, we obtain
\be
\left[ \int_0^X T_{\{xt\}} \, dx \right]_0^T
+\left[ \int_0^T T_{\{xx\}} \, dt \right]_0^X = 0 \,.
\label{intEM}
\ee
Assume now that $\phi$ is strictly periodic in $t$ with period
$T$. Then the first term above vanishes and as $X$ is arbitrary,
eq.(\ref{intEM}) simplifies to the statement that the momentum flux
integrated over a period,
\be
\int_0^T T_{\{xx\}} \, dt \,,
\label{intMom}
\ee
is independent of $x$. If, additionally, the field $\phi$ is
spatially localised, i.e. $T_{\{xx\}}$ decays to zero sufficiently fast as
$|x| \to \infty$, then
\be
\int_0^T T_{\{xx\}} \, dt = 0
\label{intzero}
\ee
for all $x$ \cite{GomideGuardiaSearaZeng2025}. This is the spatial
conservation law on which we focus.

Spatially localised and strictly periodic solutions are called
``breathers'', but they are exceptional in nonlinear scalar field
theories~\cite{SegurKruskal1987,Kichenassamy1991,
GomideGuardiaSearaZeng2025}. The best-known non-trivial example occurs in
$(1+1)$-dimensions in the sine-Gordon model, with
$V(\phi)=1-\cos\phi$. For a much wider range of potentials $V$, and also
in higher-dimensional scalar field theories, there exist ``oscillon''
solutions~\cite{Gleiser1994,CopelandGleiserMuller1995,
HindmarshSalmi2006,SaffinTranberg2007,
FodorForgacsHorvathLukacs2008,MendoncaOliveira2022}.
These are localised solutions that are close to periodic if their
amplitude is small. The frequency of a small-amplitude oscillon lies
just below the mass threshold 1, and slowly approaches 1 as it
radiates energy away and its amplitude decreases. Oscillons are
closely related to quasi-breathers, which are strictly periodic but
not completely localised~\cite{FodorForgacsGrandclementRacz2006}.
The nonlinear dynamics of oscillons, and not simply their existence,
has been extensively investigated, partly numerically; see
Ref.~\cite{Fodor2019} for a review. For some more recent studies,
and references to earlier work, see
Refs.~\cite{MantonRomanczukiewicz2023,
BlaschkeRomanczukiewiczSlawinskaWereszczynski2024,
EvslinRomanczukiewiczSlawinskaWereszczynski2025,
CamposMantonMohammadi2026}.

We are interested here in the small-amplitude, asymptotic expansion
of oscillons developed by Kichenassamy \cite{Kichenassamy1991},
Fodor et al. \cite{FodorForgacsHorvathLukacs2008} and others. This is a
power series in a small amplitude parameter $\eps$, and needs to
be truncated at some finite order $N$ to provide a good approximation
to an oscillon\footnote{We refer to these truncated asymptotic expansions
approximating oscillons as ``Fodor oscillons''.}. The Fodor
oscillon $\PhiN(t,x)$ is a polynomial
of order $N$ in $\eps$ that is strictly periodic in $t$ and whose
spatial coefficients are exponentially localised in $x$. It is an
approximation in which radiation is neglected. $\PhiN(t,x)$ fails
to exactly satisfy the field equation, so an interesting
issue is how closely the integrated momentum flux (\ref{intMom})
approaches zero. We will show that it has a leading-order
dependence at order $\eps^{N+4}$ for even $N$ and
$\eps^{N+5}$ for odd $N$ (except for $N=1$, where the order
is $\eps^4$). This vanishing to high order in $\eps$ confirms
that the Fodor oscillons provide good approximations to solutions of the
field equation.

\section{Small-amplitude expansion}
\label{sec:setup}

The field equation (\ref{eq:field_equation}) can be written as
\begin{equation}
\phi_{tt} - \phi_{xx} + \phi + W(\phi) = 0
\label{Wfieldeq}
\end{equation}
where 
\be
W(\phi) = \sum_{j=2}^{\infty} g_j \phi^j
\label{Wdefn}
\ee
denotes the sum of nonlinear terms.

Following Kichenassamy and Fodor et al., we assume that $\phi(t,x)$
has an amplitude of order $\eps \ll 1$ and is periodic with period
$2\pi/\omega$, where $\omega^2=1-\eps^2$ exactly. We then introduce
rescaled time and space variables $\tau=\omega t$ and $\zeta=\epsilon x$,
in terms of which the field equation becomes
\begin{equation}
(1 - \eps^2)\phi_{\tau\tau} - \eps^2 \phi_{\zeta\zeta} + \phi + W(\phi) = 0 \,.
\label{eq:scaled_field_equation}
\end{equation}
Introducing the operators
\begin{equation}
L=\partial_{\tau\tau}+1 \,,
\qquad
A=\partial_{\tau\tau}+\partial_{\zeta\zeta} \,,
\label{eq:operators}
\end{equation}
eq.~(\ref{eq:scaled_field_equation}) takes the more compact form
\begin{equation}
L\phi -\eps^2 A\phi + W(\phi)=0 \,.
\label{eq:compact_field_equation}
\end{equation}
Next, the field $\phi$ is expanded in a power series in $\eps$,
\begin{equation}
\phi(\tau,\zeta)=\sum_{n=1}^{\infty}\eps^n \phi_n(\tau,\zeta) \,.
\label{eq:Fodor_expansion}
\end{equation}
Substituting into eq.~(\ref{eq:compact_field_equation})
and collecting terms at each order in $\eps$ gives a sequence of field
equations. Up to fifth order, these are
\bea
&{}& L\phi_1=0 \,,
\label{eq:order1}
\\
&{}& L\phi_2 + g_2\phi_1^2 = 0 \,,
\label{eq:order2}
\\
&{}& L\phi_3 -A\phi_1 + 2g_2\phi_1\phi_2 + g_3\phi_1^3 = 0 \,,
\label{eq:order3}
\\
&{}& L\phi_4 - A\phi_2 +g_2(2\phi_1\phi_3+\phi_2^2)
+ 3g_3\phi_1^2\phi_2 + g_4\phi_1^4 = 0 \,,
\label{eq:order4}
\\
&{}& L\phi_5 - A\phi_3 + 2g_2(\phi_1\phi_4 + \phi_2\phi_3)
+ 3g_3 (\phi_1^2\phi_3 + \phi_1\phi_2^2) + 4g_4
\phi_1^3\phi_2 + g_5\phi_1^5 = 0 \,.
\label{eq:order5}
\eea
This is effectively a sequence of ODEs in $\tau$ that can be solved
iteratively. The coupling $g_j$ first enters the equations at order $\eps^j$.

It is consistent to assume that $\phi_n$ is a
finite combination of Fourier harmonics $\cos k\tau$ with the parity of $k$
equal to that of $n$, i.e. $k$ even (odd) if $n$
is even (odd). These harmonics are multiplied by localised
functions of $\zeta$. The source terms for $L\phi_n$ are a
combination of products of harmonics arising from
$\phi_1,\dots,\phi_{n-1}$, all with the parity of $n$, and a solution
can be constructed using the particular integrals 
\be
L\chi = \cos k\tau \ \Longrightarrow \
\chi(\tau) = \frac{1}{1-k^2} \cos k\tau
\ee
for $k \ne 1$. There can also be (and is) a homogeneous contribution
to $\phi_n$ proportional to $\cos\tau$ when $n$ is 
odd, but not when $n$ is even, as it would spoil the parity of $\phi_n$.
The non-trivial step in the iteration is to fix the homogeneous part of
$\phi_n$ for $n$ odd, so as to ensure that the source for
$L\phi_{n+2}$ has no term proportional to $\cos\tau$, as that would
lead to resonance and lack of periodicity in $\phi_{n+2}$. There is
no such constraint for $n$ even, as the source automatically
has no term proportional to $\cos\tau$. It is the non-resonance condition
that determines the $\zeta$-dependence of the Fodor oscillon
up to any finite order. 

The solution of the first- and second-order equations
(\ref{eq:order1}) and (\ref{eq:order2}) is
\begin{equation}
\phi_1(\tau,\zeta)= p_1(\zeta)\cos\tau \,, \quad
\phi_2(\tau,\zeta)=-\frac{1}{6} g_2 \, p_1^2(\zeta)
(3-\cos 2\tau) \,,
\label{eq:phi12}
\end{equation}
where the leading-order spatial coefficient $p_1(\zeta)$ is determined
by the non-resonance condition in the third-order equation
(\ref{eq:order3}), namely
\begin{equation}
p_1'' - p_1 + \lambda \, p_1^3=0 \,, \quad {\rm where} \quad
\lambda=\frac{5}{6}g_2^2 - \frac{3}{4}g_3 \,.
\label{eq:p1_equation}
\end{equation}
For $\lambda>0$, the spatially localised solution is
\begin{equation}
p_1(\zeta)=\sqrt{\frac{2}{\lambda}} \,
\operatorname{sech}\zeta \,.
\label{eq:p1_solution}
\end{equation}
This construction does not give a localised oscillon solution if
$\lambda \leq 0$, i.e. if the quartic term in the potential $V$ is too large. 

With $p_1$ determined, the third- and fourth-order equations
(\ref{eq:order3}) and (\ref{eq:order4}) can be solved. In particular,
$\phi_3$ has the form
\begin{equation}
\phi_3(\tau,\zeta) = p_3(\zeta)\cos\tau
+\frac{1}{96} (2g_2^2 + 3g_3) \, p_1^3(\zeta)\cos 3\tau \,,
\label{eq:phi3}
\end{equation}
with the homogeneous coefficient $p_3(\zeta)$ determined by the
non-resonance condition in equation (\ref{eq:order5}) for $\phi_5$. The
iteration can be continued indefinitely: In $\phi_n(\tau,\zeta)$,
polynomials in $\sech \zeta$ multiply a finite set of harmonics
$\cos k\tau$. The explicit solution up to fourth order is given for
the simplest, cubic oscillon \cite{MantonRomanczukiewicz2023}
with $g_2 = -1$ and $g_j = 0 \ (j \geq 3)$ in section V. 

\section{Evaluating the integrated momentum flux}

In terms of the rescaled variables, the integrated momentum flux
(\ref{intMom}) becomes
\begin{align}
  Q = \frac{1}{\omega} \intd \left( \half \omega^2 \phi^2_{\tau}
+ \half \eps^2 \phi^2_{\zeta} - V(\phi)\right) d\tau \,.  
\label{mom-flux-int}
\end{align}
It is easier to analyse the $\zeta$-derivative of $Q$, which involves
the field equation (\ref{eq:compact_field_equation}). One finds
\be
Q_{\zeta} = -\frac{1}{\omega} \int_0^{2\pi} \phi_{\zeta}
\left( L\phi -\eps^2 A\phi + W(\phi) \right) d\tau
\label{IntFieldeq}
\ee
after integrating one of the terms by parts. This last step requires
$\phi$ to be periodic, but the field equation does not need to be satisfied.

Introduce the notation
\be
F = L\phi -\eps^2 A\phi + W(\phi)
\ee
for the left-hand side of eq.(\ref{eq:compact_field_equation}).
$F$, like $\phi$, has an expansion in powers of $\eps$,
\be
F = \eps F_1 + \eps^2 F_2 + \cdots + \eps^n F_n + \cdots \,. 
\ee
We refer to $F_i = 0$ as the $i$th-order field equation. Up to fifth order
these are eqs. (\ref{eq:order1}) -- (\ref{eq:order5}).

We wish to calculate $Q_\zeta$ for the truncated, Fodor oscillon
\be
\PhiN = \eps \phi_1 + \cdots + \eps^N \phi_N \,.
\ee
$\PhiN$ satisfies $F_i = 0$ for $i \leq N$, but fails to satisfy the
higher-order equations. Therefore, when $\PhiN$ is substituted into
the complete field equation, we obtain a series
\be
F = \eps^{N+1} R_{N+1} + \eps^{N+2} R_{N+2} + \cdots \,,
\ee
where $R_{N+i}$ is referred to as the $(N+i)$th-order residual. It contains
all the terms of $F_{N+i}$ that involve only $\phi_1 \,, \dots \,, \phi_N$.
For example, for the first-order truncation $\Phi^{\{ 1 \}}$ the
residual $R_3$ is $ -A\phi_1 + g_3 \phi_1^3$.
Using this notation,
\be
Q_\zeta = -\frac{1}{\omega}\intd
\left(\eps \phi_{1,\zeta} + \cdots + \eps^N \phi_{N,\zeta}\right)
\times
\left(\eps^{N+1} R_{N+1} + \eps^{N+2} R_{N+2} +
  \eps^{N+3} R_{N+3} + \cdots \right) d\tau \,.
\label{Presid}
\ee
This can be reduced to a function of $\zeta$ by multiple use of the
orthogonality relation for Fourier harmonics
\be
\intd \cos k\tau \, \cos l\tau \, d\tau
= \pi \, \delta_{kl} \,,
\ee
valid if $k$ and $l$ are not both 0, together with the value
$2\pi$ if $k=l=0$.

Also recall that the terms at order $\eps^n$ in either bracket of
(\ref{Presid}) are combinations of harmonics with the same parity
as $n$. As the integral of a product of harmonics with opposite parity
vanishes, all nonzero terms of $Q_\zeta$ are of even order in $\eps$. 

Working with residuals is practical for truncations at very low order, but
becomes cumbersome at higher order. It is therefore helpful to
introduce the notion of a dual residual $S_{N+i}$, defined by
\be
F_{N+i} = R_{N+i} + S_{N+i} \,,
\ee
where $F_{N+i} = 0$ is the complete $(N+i)$th-order field equation.
$S_{N+i}$ contains all the terms in $F_{N+i}$ that have at least
one contribution from $\phi_{N+j}$ with $j > 0$. For example, for the
truncation $\Phi^{\{ 1 \}}$ the dual residual $S_3$ is
$L \phi_3 + 2g_2\phi_1\phi_2$. The dual residuals for $i \leq 3$ are
all that we will need and these have relatively few terms.
To exploit the dual residuals, we imagine that, although the
truncation is at order $N$, we have solved for the higher-order terms,
so that $F_{N+i} = 0$, at least for $i \leq 3$. In the integral
$Q_\zeta$ we can then replace $R_{N+i}$ by $-S_{N+i}$ and obtain
\be
Q_\zeta = \frac{1}{\omega} \intd
\left(\eps \phi_{1,\zeta} + \cdots + \eps^N \phi_{N,\zeta}\right)
\times
\left( \eps^{N+1} S_{N+1} + \eps^{N+2} S_{N+2} +
  \eps^{N+3} S_{N+3} + \cdots \right) d\tau \,.
\label{Pdual}
\ee

We will present in the following section the form taken by
$Q_\zeta$ at leading order in $\eps$. This is shown to be a total
derivative with respect to $\zeta$. Therefore, we immediately obtain
an expression for $Q$ (which is still an integral over a
period $2\pi$ in $\tau$). This expression has special features
for the low-order truncations $N=1$, 2 and 3, but a pattern emerges
for $N \geq 4$, depending on the parity of $N$.

\section{The integrated momentum flux at leading order}
\label{sec:general_truncations}

$\PhiN$, the $N$th-order truncation, satisfies the field
equations up to order $N$, but we need to assume that solutions
of a few of the next-higher-order field equations have been
determined. So we suppose that the iteration has been continued
indefinitely, and write
\be
\phi = \PhiN + \psi
\ee
where
\be
\psi = \eps^{N+1} \phi_{N+1} + \eps^{N+2} \phi_{N+2} + \eps^{N+3}
\phi_{N+3} + \cdots \,.
\ee
The complete field equation is formally
\be
L(\PhiN + \psi) - \eps^2 A(\PhiN + \psi) + W(\PhiN + \psi) = 0
\ee
where $L$ and $A$ act linearly. We expand this out as
\be
L\PhiN -\eps^2 A\PhiN + W(\PhiN)
+ L\psi -\eps^2 A\psi + W'(\PhiN)\psi + O(\psi^2) = 0
\label{fieldeqexpand}
\ee
and neglect the $O(\psi^2)$ contribution for now.
As the field equation is satisfied by $\PhiN$ up to order $N$,
the combined terms $L\PhiN -\eps^2 A\PhiN + W(\PhiN)$ contain
all the residuals $R_{N+i}$. $\psi$ contains the fields
$\phi_{N+j}$, so $L\psi -\eps^2 A\psi + W'(\PhiN)\psi$ contains all
the dual residuals $S_{N+i}$ at linear order in $\psi$. 

From the expression (\ref{Pdual}) we deduce that
\begin{align}
Q_\zeta= \frac{1}{\omega} \intd \PhiN_\zeta \left( L\psi -\eps^2 A\psi
+ W'(\PhiN)\psi \right) d\tau,
\label{eq:P_general_1}
\end{align}
and the self-adjointness of $L$ implies that $L$ can be shifted to act on
$\PhiN_\zeta$. From eq.(\ref{fieldeqexpand}) we have 
\begin{equation}
L\PhiN = \eps^2 A\PhiN - W(\PhiN) - L\psi \,,
\end{equation}
where the terms $\eps^2 A\psi - W'(\PhiN)\psi$ and beyond can all be
neglected, as they are subleading compared to $L\psi$. The term
$L\psi$ is also generally subleading, but makes an important
contribution for small $N$. Differentiating with respect to $\zeta$
then gives
\begin{align}
L \PhiN_\zeta = \eps^2 A\PhiN_\zeta - W'(\PhiN) \PhiN_\zeta - L\psi_\zeta.
\label{eq:Lphi_derivative}
\end{align}
Note that the terms on the right-hand side generally have different
orders in $\eps$, and we will clarify below which terms dominate and
which can be dropped. The key observation is that $W'(\PhiN)$ appears
both here and in the linear term of the expansion of
$W(\PhiN + \psi)$. Substituting Eq.~(\ref{eq:Lphi_derivative}) into
Eq.~(\ref{eq:P_general_1}) therefore gives
\begin{equation}
Q_\zeta = \frac{1}{\omega}
\intd \left( \eps^2 (A\PhiN_\zeta) \psi
- W'(\PhiN) \PhiN_\zeta \psi - \psi L\psi_\zeta - \eps^2 \PhiN_\zeta A\psi 
+ \PhiN_\zeta W'(\PhiN) \psi \right) d\tau \,,
\end{equation}
where the second and final terms cancel exactly, and we are left with
\begin{equation}
Q_\zeta = \frac{1}{\omega} 
\intd \left( \eps^2 (A\PhiN_\zeta) \psi - \eps^2 \PhiN_\zeta A\psi 
- \psi L\psi_\zeta \right) d\tau \,.
\label{eq:P_A_only}
\end{equation}
All explicit dependence on the couplings $g_j$ disappears, although it remains
implicit in $\PhiN$. 

Recall that  $A=\pr_{\tau\tau}+\pr_{\zeta\zeta}$, and note that the
terms in (\ref{eq:P_A_only}) containing $\pr_{\tau\tau}$ cancel by
self-adjointness. Also, for functions $f$ and $g$,
\be
f_{\zeta\zeta\zeta} \, g - f_\zeta \, g_{\zeta\zeta}
=\partial_\zeta (f_{\zeta\zeta} \, g - f_\zeta \, g_\zeta) \,.
\label{eq:A_identity}
\ee
Further, note that $L$ commutes with the $\zeta$-derivative.
Therefore, the above expression for $Q_\zeta$ is a total  $\zeta$-derivative
\begin{equation}
Q_\zeta =  \frac{1}{\omega} \frac{d}{d\zeta}
\intd \left( \eps^2 ( \PhiN_{\zeta\zeta} \psi - \PhiN_\zeta \psi_\zeta )
- \half \psi L \psi \right) d\tau
\label{eq:master_relation}
\end{equation}
and $\zeta$-integration gives
\begin{equation}
Q =  \frac{1}{\omega} \intd
\left( \eps^2 (\PhiN_{\zeta\zeta} \psi - \PhiN_\zeta \psi_\zeta)
- \half \psi L \psi \right) d\tau \,,
\label{eq:masterP}
\end{equation}
since $Q$ vanishes at spatial infinity.

From this we can obtain our final, leading-order expressions for $Q$.
The result is different for even and odd $N$, so we present these
cases separately. 

\subsection{Even truncations}

Let $N = 2M$. Then
\begin{equation}
\PhiN = \sum_{n=1}^{2M} \eps^n \phi_n
\end{equation}
and
\begin{equation}
\psi = \eps^{2M+1}\phi_{2M+1} + \eps^{2M+2}\phi_{2M+2}+\cdots.
\end{equation}
In $Q$, the leading contribution comes from the pair
$(\phi_1, \phi_{2M+1})$, and is
\begin{equation}
Q = \frac{1}{\omega}
\intd \left( \eps^{2M+4} \left( \phi_{1,\zeta\zeta} \, \phi_{2M+1} - 
\phi_{1,\zeta}\, \phi_{2M+1,\zeta} \right) - \half \eps^{4M+2} \phi_{2M+1} L
\phi_{2M+1} \right) d\tau \,.
\label{eq:P_general_even}
\end{equation}
For $M>1$ we have $2M+4 < 4M+2$ so the final term is subleading and can
be dropped. Then, as $\phi_1(\tau,\zeta) = p_1(\zeta) \cos\tau$, and the
only part of $\phi_{2M+1}$ that is proportional to $\cos\tau$ is the
homogeneous term $p_{2M+1}(\zeta) \cos\tau$, $Q$ simplifies after
integrating over $\tau$ to
\be
Q = \frac{\pi}{\omega} \, \eps^{2M+4} \,
(p''_1 \, p_{2M+1} - p'_1 \, p'_{2M+1}) \,.
\label{Qevenexplicit}
\ee
This is a general formula but not easy to evaluate explicitly,
as $p_{2M+1}(\zeta)$ needs to be determined from the equation
imposing non-resonance for $\phi_{2M+3}$.

The truncation $\Phi^{\{ 2 \}}$ with $M=1$ is special, as $2M+4 =
4M+2 = 6$ and we need to keep all terms in (\ref{eq:P_general_even}). So in
this case, at leading order,
\begin{equation}
Q = \frac{1}{\omega} \eps^6 
\intd \left( (\phi_{1,\zeta\zeta} \, \phi_3 - \phi_{1,\zeta}\, \phi_{3,\zeta})
- \half \phi_3 L \phi_3 \right) d\tau \,.
\label{Qorder2}
\end{equation}
Using the expressions (\ref{eq:phi12}) and (\ref{eq:phi3}) for
$\phi_1$ and $\phi_3$, and integrating over $\tau$, this simplifies to
\be
Q = \frac{\pi}{\omega} \eps^6 \left( p''_1 \, p_3 - p'_1 \, p'_3 +
  \frac{1}{2304} (2g_2^2 + 3g_3)^2 p_1^6 \right) \,.
\ee

\subsection{Odd truncations}

Here, $N=2M-1$, so
\begin{equation}
\PhiN = \sum_{n=1}^{2M-1} \eps^n\phi_n
\end{equation}
and
\begin{equation}
\psi=\eps^{2M} \phi_{2M} +\eps^{2M+1} \phi_{2M+1}+\cdots \,.
\end{equation}
By the parity argument, $Q$ has terms only at even order in $\eps$, and
the leading term has contributions from both the pairs
$(\phi_1,\phi_{2M+1})$ and $(\phi_2,\phi_{2M})$. 

Steps similar to those leading to eq.(\ref{eq:P_general_even}) give
\bea
Q &=& \frac{1}{\omega} 
\intd \bigg( \eps^{2M+4} \left( \phi_{1,\zeta\zeta} \, \phi_{2M+1} - 
\phi_{1,\zeta} \, \phi_{2M+1,\zeta} \right)
+ \eps^{2M+4} \left( \phi_{2,\zeta\zeta} \, \phi_{2M}
- \phi_{2,\zeta} \, \phi_{2M,\zeta} \right) \nonumber \\
&& \qquad \qquad - \half \eps^{4M} \phi_{2M} L \phi_{2M} \bigg) d\tau \,.
\label{eq:P_general_odd}
\eea
For $M>2$ we have $2M+4 < 4M$ so the last term is subleading and can
be dropped. Then, as before, $\phi_1$ only involves $\cos\tau$, so
only the homogeneous part of $\phi_{2M+1}$ contributes. On the other
hand, $\phi_2$ combines the harmonics $1$ and $\cos 2\tau$, so these
same harmonics in $\phi_{2M}$ are required to complete the integration
over $\tau$. 

Among the odd truncations, $\Phi^{\{ 3 \}}$ and $\Phi^{\{ 1 \}}$ are special.
For the truncation $\Phi^{\{ 3 \}}$ with $M=2$, we have $2M+4 =
4M = 8$ and need to keep all terms in (\ref{eq:P_general_odd}). In
this case
\begin{align}
Q = \frac{1}{\omega} \eps^8  
\intd \left( \left( \phi_{1,\zeta\zeta} \, \phi_5 - 
\phi_{1,\zeta} \, \phi_{5,\zeta} \right)
+ \left( \phi_{2,\zeta\zeta} \, \phi_4 - \phi_{2,\zeta} \, \phi_{4,\zeta}
\right) - \half \phi_4 L \phi_4 \right) d\tau \,.
\end{align}
which is more tricky than (\ref{Qorder2}) to simplify.

For the truncation $\Phi^{\{ 1 \}}$ with $M=1$, we have $2M+4 > 4M$ and,
unusually, need to keep just the last term in
(\ref{eq:P_general_odd}), so
\begin{align}
Q = -\frac{1}{2\omega} \eps^4 \intd \phi_2 L \phi_2 \, d\tau \,.
\end{align}
Explicitly, using the solution (\ref{eq:phi12}) for $\phi_2$, we have
\be
\phi_2 = -\frac{1}{6} g_2 \, p_1^2(\zeta) (3 - \cos 2\tau) \,, \quad
L\phi_2 = -\frac{1}{2} g_2 \, p_1^2(\zeta) (1 + \cos 2\tau)
\ee
so
\be
Q = -\frac{5\pi}{24\omega} \eps^4 \, g_2^2 \, p_1^4(\zeta) \quad {\rm where}
\quad p_1(\zeta)=\sqrt{\frac{2}{\lambda}} \,
\operatorname{sech}\zeta \,.  
\ee
This is of order $\eps^4$, rather than order $\eps^6$ as would be
anticipated from the general result for odd-order truncations.

In summary, the truncations at even order $N=2M$ and at odd order $N=2M-1$
almost always result in leading contributions to $Q$ at the same order,
$\eps^{2M+4}$, i.e. $\eps^{N+4}$ for even $N$ and $\eps^{N+5}$ for odd
$N$. The formulae for $Q$ contain the same term involving
$\phi_1$ and $\phi_{2M+1}$, while the formula for odd $N$ contains
an additional term involving $\phi_2$ and $\phi_{2M}$. Exceptionally,
for the first-order truncation, $Q$ has order $\eps^4$.

In these calculations, we have neglected possible contributions to
$Q$ from $\PhiN_\zeta \psi^2$ and $\eps^2(A\psi)\psi$. $\psi^2$ has
even parity, so the leading relevant term in $\PhiN$ is
$\eps^2\phi_2$. $\psi^2$ has leading order $\eps^{2N+2}$, so the
total order of both these contributions is $\eps^{2N+4}$, which is
subleading for all $N$. 

\section{Example: the cubic oscillon}
\label{sec:example}

The general results above are illustrated by the oscillon for the simplest,
cubic potential
\begin{equation}
V(\phi)=\frac{1}{2}\phi^2-\frac{1}{3}\phi^3,
\end{equation}
which satisfies the field equation
\begin{equation}
\phi_{tt}-\phi_{xx}+\phi-\phi^2=0 \,.
\label{eq:example_field_equation}
\end{equation}
Here, $g_2=-1$ and $g_j=0$ for $j\geq3$, so
$\lambda=\frac{5}{6}$, and the oscillon spatial profile at first order is
\begin{equation}
p_1(\zeta)
=
\sqrt{\frac{12}{5}}\,\operatorname{sech}\zeta \,.
\label{eq:example_p1}
\end{equation}
The function $p_3(\zeta)$, determined by the non-resonance condition, is
\begin{equation}
p_3(\zeta)
=
\frac{47\sqrt{15}}{75}\operatorname{sech}\zeta
-\frac{119\sqrt{15}}{150}\operatorname{sech}^3\zeta \,.
\label{eq:example_p3}
\end{equation}

\begin{figure}[t]
\centering
\includegraphics[width=\columnwidth]{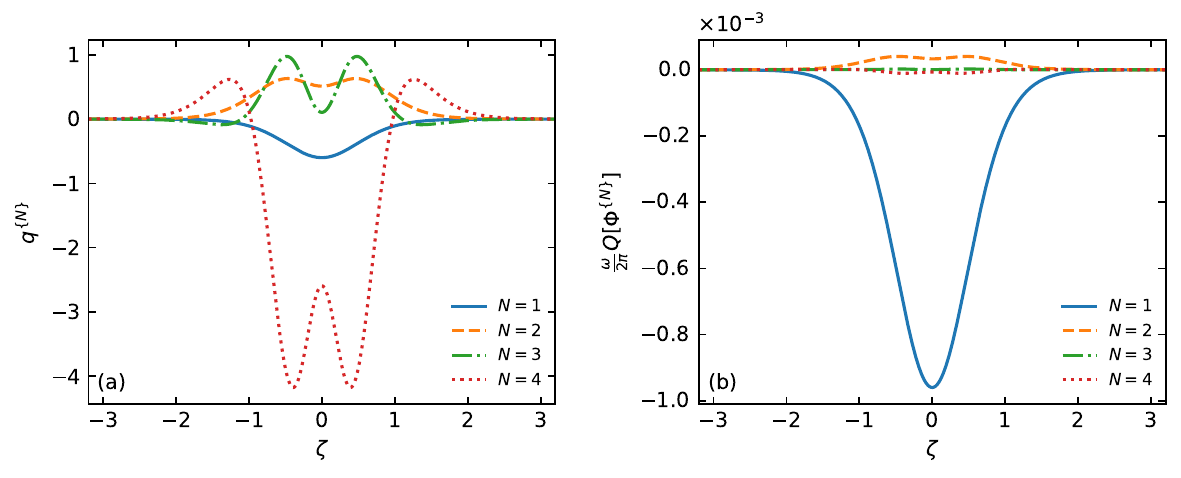}
\caption{Integrated momentum flux for the cubic oscillon: 
(a) Normalised spatial dependence $q^{\{N\}}$, 
(b) Spatial dependence of $Q[\Phi^{\{N\}}]$ for $\epsilon=0.2$.}
\label{Q_cubic}
\end{figure}

The first four terms in the Fodor expansion are therefore
\cite{MantonRomanczukiewicz2023}
\begin{align}
\phi_1(\tau,\zeta)
&=
\sqrt{\frac{12}{5}}\,
\operatorname{sech}\zeta\cos\tau \,,
\label{eq:example_phi1}
\\
\phi_2(\tau,\zeta)
&=
\frac{2}{5}\,
\operatorname{sech}^2\zeta
\left(3-\cos2\tau\right) \,,
\label{eq:example_phi2}
\\
\phi_3(\tau,\zeta)
&=
\left(
\frac{47\sqrt{15}}{75}\operatorname{sech}\zeta
-\frac{119\sqrt{15}}{150}\operatorname{sech}^3\zeta
\right)\cos\tau
+\frac{\sqrt{15}}{50}
\operatorname{sech}^3\zeta\cos3\tau \,,
\label{eq:example_phi3}
\\
\phi_4(\tau,\zeta)
&=
\frac{214}{25}\operatorname{sech}^2\zeta
-\frac{261}{25}\operatorname{sech}^4\zeta
\nonumber \\
& \quad +\left(
-\frac{94}{75}\operatorname{sech}^2\zeta
+\frac{16}{15}\operatorname{sech}^4\zeta
\right)\cos2\tau
-\frac{1}{75}\operatorname{sech}^4\zeta\cos4\tau \,.
\label{eq:example_phi4}
\end{align}
The leading contributions to the integrated momentum flux $Q$ for the
first- to fourth-order truncations are calculated to be
\begin{align}
Q[\Phi^{\{1\}}]
&=
\frac{2\pi}{\omega}\eps^4
\left[
-\frac{3}{5}\operatorname{sech}^4\zeta
\right] ,
\\
Q[\Phi^{\{2\}}]
&=
\frac{2\pi}{\omega}\eps^6
\left[
\frac{72}{25}\operatorname{sech}^4\zeta
-\frac{296}{125}\operatorname{sech}^6\zeta
\right] ,
\\
Q[\Phi^{\{3\}}]
&=
\frac{2\pi}{\omega}\eps^8
\left[
-\frac{8977}{1500}\operatorname{sech}^4\zeta
+\frac{56431}{2500}\operatorname{sech}^6\zeta
-\frac{61813}{3750}\operatorname{sech}^8\zeta
\right] ,
\\
Q[\Phi^{\{4\}}]
&=
\frac{\pi}{\omega}\eps^8
\left[
\frac{14737}{250}\operatorname{sech}^4\zeta
-\frac{142063}{750}\operatorname{sech}^6\zeta
+\frac{31321}{250}\operatorname{sech}^8\zeta
\right] .
\label{eq:example-res}
\end{align}
These expressions may be obtained either from the general formulae of
Section~\ref{sec:general_truncations} or by direct substitution of
formulae (\ref{eq:example_phi1})--(\ref{eq:example_phi4}) into
(\ref{mom-flux-int}). In the left panel of Fig.~\ref{Q_cubic}, we
compare the normalised spatial dependences $q^{\{N\}}$ of
$Q[\Phi^{\{N\}}]$ for the first- to fourth-order truncations, defined by
\begin{equation}
Q[\Phi^{\{N\}}](\zeta)=\frac{2\pi}{\omega}\eps^{p_N}q^{\{N\}}(\zeta) \,,
\quad {\rm with} \quad
p_1=4,\quad p_2=6,\quad p_3=p_4=8 \,.
\label{eq:qN-definition}
\end{equation}
In the right panel, we compare $Q[\Phi^{\{N\}}]$ itself
for the fixed value $\eps=0.2$, with $\omega=\sqrt{1-\eps^2}$. The
figure confirms the smallness of its deviation from zero for the
higher-order Fodor oscillons $\Phi^{\{3\}}$ and $\Phi^{\{4\}}$.

\section{Conclusions}
\label{sec:conc}

We have analysed the integrated momentum flux $Q$ for truncated, Fodor
oscillons. The Fodor oscillon is exactly periodic, but does not
solve the field equation exactly. So $Q$, which would vanish for an
exact breather, is here non-zero. Although the first omitted term in the
$N$th-order truncation is $O(\eps^{N+1})$, where $\eps$ is the
small amplitude parameter, $Q$ is considerably smaller, being
$O(\eps^{N+4})$ for even $N$ and $O(\eps^{N+5})$ for odd $N$,
except for $N=1$, where it is $O(\eps^4)$. Strikingly, for the
third- and fourth-order truncations, which are useful for oscillons
of modest amplitude, $Q$ is $O(\eps^8)$, and therefore very
small. We have illustrated the spatial dependence of $Q$ 
for the simplest, cubic oscillon.

The momentum flux could also be used to improve approximate oscillon
profiles. For example, adjustable parameters in the spatial profile
could be chosen to make $Q$ as small as possible. In this way, $Q$
could provide a simple measure of the accuracy of an approximate
oscillon, complementary to the residual of the field equation. This
may be particularly useful for oscillons of moderate amplitude, where
the truncated, small-amplitude expansion is less accurate.

Similar calculations should be possible for kinks excited by their
shape modes. These are periodic configurations that do not exactly
solve their nonlinear field equations. Extensions of the results here to
oscillons in higher spatial dimensions could also be interesting. 

\vskip 200pt
\section*{Acknowledgements}

NSM was partially supported by the STFC Consolidated Grant ST/P000681/1.
AM acknowledges financial support from CNPq (Conselho Nacional de
Desenvolvimento Científico e Tecnológico), Grant No.~306295/2023-7,
and from CAPES (Coordenação de Aperfeiçoamento de Pessoal de Nível
Superior).

\end{document}